\documentstyle[12pt]{article}

\newcommand{\nc}{\newcommand}
\nc{\beq}{\begin{equation}}
\nc{\eeq}{\end{equation}}
\nc{\beqa}{\begin{eqnarray}}
\nc{\eeqa}{\end{eqnarray}}

\input{epsf.sty}  
\input{psfig.sty}
\newwrite\ffile\global\newcount\figno \global\figno=1

\def\writedef#1{}
\def\figin{\epsfcheck\figin}\def\figins{\epsfcheck\figins}
\def\epsfcheck{\ifx\epsfbox\UnDeFiNeD
\message{(NO epsf.tex, FIGURES WILL BE IGNORED)}
\gdef\figin##1{\vskip2in}\gdef\figins##1{\hskip.5in}
\else\message{(FIGURES WILL BE INCLUDED)}%
\gdef\figin##1{##1}\gdef\figins##1{##1}\fi}
\def\figinsert{}
\def\ifig#1#2#3{\xdef#1{fig.~\the\figno}
\writedef{#1\leftbracket fig.\noexpand~\the\figno}%
\figinsert\figin{\centerline{#3}}\medskip\centerline{\vbox{\baselineskip12pt
\advance\hsize by -1truein\center\footnotesize{  Fig.~\the\figno.} #2}}
\bigskip\endinsert\global\advance\figno by1}
\def\endinsert{}

\begin{document}

\title{\large{\bf 
Flavour Universal 
Dynamical Electroweak Symmetry Breaking}}

\author{
Gustavo Burdman\thanks{burdman@pheno.physics.wisc.edu} \\
{\small\em Department of Physics, University of Wisconsin, Madison, 
WI 53706.} \\ \\
Nick Evans\thanks{nevans@budoe.bu.edu}
 \\ {\small\em Department of Physics,
Boston University, Boston, MA 02215.} \\ 
}

\date{\mbox{  }}

\maketitle

\begin{picture}(0,0)(0,0)
\put(350,310){BUHEP-98-28}
\put(350,295){MADPH-98-1075}
\end{picture}
\vspace{-24pt}

\vspace{-0.6cm}

\begin{abstract}
The top condensate see-saw mechanism of Dobrescu and Hill allows
electroweak symmetry to be broken while deferring the problem of
flavour to an electroweak singlet, massive sector. We provide an
extended version of the singlet sector that naturally accommodates
realistic masses for all the standard model fermions, 
which play an equal role in
breaking electroweak symmetry. The models result in a relatively light
composite Higgs sector with masses typically in the range of 
(400-700)~GeV. In more
complete models the dynamics will presumably 
be driven by a broken gauged family or flavour symmetry group.
As an example of the higher scale dynamics
a fully dynamical model of the quark sector with a GIM mechanism
is presented, based on an earlier top condensation model of King
using broken family gauge symmetry interactions
(that model
was itself based on a technicolour model of Georgi). 
The crucial extra
ingredient is a reinterpretation of the condensates that form when 
several gauge groups become strong close to the same scale. A related
technicolour model of Randall which naturally includes the leptons too
may also be adapted to this scenario. We discuss the 
low energy constraints on the massive gauge bosons and scalars
of these models as well as their  
phenomenology at the TeV scale.
\end{abstract}

\newpage

\section{Introduction}
  
The electroweak symmetry (EWS) of the standard model, which is a chiral 
symmetry of the fermions, is spontaneously 
broken by some as yet unknown mechanism. The introduction of an elementary
Higgs scalar is technically unnatural. A natural explanation for the
breaking of the chiral symmetry is a higher scale repeat of the dynamical 
breaking induced by QCD. This idea spawned technicolour models \cite{TC} of EWS
breaking but these models ran into trouble since they typically introduce many
extra electroweak doublet fermions whose presence is naively in conflict
with precision experimental data from LEP and SLC \cite{peskin}. 
With the discovery of the 
large top mass it was proposed that a top quark condensate
\cite{topcond}, 
generated
by some chiral symmetry breaking but not confining interaction (most
likely a gauge group broken close to its critical scale for chiral
symmetry breaking) might be 
responsible for EWS breaking. However, the value of the top quark mass
is too small to generate
the EWS breaking scale $v$. 
Introducing extra fermions in this fashion
again leads to conflict with electroweak
precision measurements. 
A further problem, also
found in technicolour, is that attempts to generate the top-bottom mass
splitting typically give rise to large custodial isospin violating effects
in the massive sector \cite{cdt}. Such effects are harshly constrained by the 
precision measurements of the $\rho$ parameter.
A possible alternative is the combination of top condensation as the source of a 
large dynamical top quark mass, and technicolour as giving most of the EWS breaking. 
In this scenario, known as topcolor assisted technicolor~\cite{tcatc},  
the extended technicolour interactions give the top quark only a small 
fraction of its mass. On the other hand, it is still necessary to have a 
large number of techni-doublets in order to obtain the correct pattern of 
light quark masses. 
A variety of models of this type as well as their 
phenomenological consequences, have been studied in the literature~\cite{tcatc2}.

Recently Dobrescu and Hill \cite{bh} 
have proposed a model in which a dynamically
generated mass for the  top quark does
generate the full electroweak symmetry breaking scale but the 
correct 
top mass is obtained as a 
result of a see-saw mechanism with a heavy fermion sector. Perhaps of more
significance, the model allows the origin of custodial symmetry breaking
to be deferred to an EWS singlet sector where it cannot contaminate the
$\rho$ parameter. 
In their model the top quark is treated in a unique fashion, its condensation 
being driven by a broken colour interaction 
unique to the third generation quarks. This essential non-universality 
disrupts the $SU(3)$ family symmetry
of the standard model (SM) in the absence of fermion masses. 
As a consequence, it is not obvious how to feed the top condensate down 
to provide masses for the leptons and lighter first and second generation quarks. 

In this paper we propose a mechanism analogous to  
the top condensate see-saw but which can
naturally accommodate all the fermion masses and generation structure. 
An interesting  aspect of the resulting models is that all the 
SM electroweak doublets
play an equal role in breaking EWS. The result of this universal
involvement in EWS breaking is that the models give rise to a relatively
light\footnote{We refer to Higgs masses in the few hundred GeV range as light, 
compared to the unitarity bound of approximately $\simeq 1.2~$TeV. A large class
of models of a strongly coupled EWS breaking sector saturate this bound. 
}
composite Higgs sector with a typical mass scale of order (400-700)~GeV.
The SM fermion mass splittings are the result 
of mass terms in an EWS singlet 
sector, which at this stage of model building is not explained dynamically but 
simply put in by hand. 
In Section~2 we present the simplest version of the model in 
which the dynamical symmetry breaking is driven by strong four fermion 
interactions. Depending on the flavour structure of the four fermion
interactions driving EWS breaking, different patterns of pseudo-Nambu-Goldstone
bosons (pNGB) result. We discuss in detail one possible pattern in
Section~3  which is analogous to the spectrum of a one family
technicolour model, and
another one in the model of Section~4 where there are no un-eaten pNGB at the
weak scale.

The simple model of Section~2 is in fact inspired by a more complete
model of the dynamics of
top condensation of King \cite{king}, 
which in turn was derived from a technicolour model by Georgi~\cite{georgi}. 
In Ref.~\cite{king}
the four fermion interactions are the result of a broken gauged family 
symmetry. The biggest success of this class of models is that they have
a GIM mechanism above the weak scale which protects them from flavour
changing neutral currents (FCNC) even with family gauge bosons with
masses of order a TeV. This is achieved by gauging the full
chiral family symmetry, the symmetry responsible for the SM
GIM mechanism. The model does have 
a number of possible flaws, including a
non-trivial assumption about vacuum alignment and potentially light
pNGB coming from the singlet sector. Nevertheless it 
is useful to elucidate the
idea of a universal see-saw. The model is also suggestive of the gauge
structure that is likely to underlie the universal see-saw model
suggesting experimental searches. 
In Section~3 we revive the model by reinterpreting the dynamics
when several gauge groups become strongly interacting at the
same scale. The model then produces the flavour universal see-saw masses.
The full model has additional interesting dynamics as a result of the 
massive, strongly interacting flavourons of the family gauge symmetry group
and colourons analogous to those of the model of Ref.~\cite{colorons}. 
Their phenomenology, together with that of the scalar and
pseudo-scalar sector originating
from the breaking of the fermionic chiral symmetries, is discussed.  
The inclusion of  leptons in that model is non-trivial. A related technicolour
model of Randall \cite{randall}, which more naturally includes the
leptons, may be adapted to provide a low energy flavour
universal EWS breaking model where the strong four fermion interactions
result from a larger gauged flavour symmetry. We discuss this model,
its flaws  and its phenomenology in Section~4.

Finally, we must address the matter of fine tuning in top condensate
models. In these models we will assume that the interactions responsible
for EWS breaking are broken gauge interactions with gauge bosons with
masses in the range $\Lambda =(1-10)~$TeV.  
The fact that their interactions give rise to the
weak scale $v=246$~GeV naively suggests fine tuning of order 
$(1-10)\%$. In fact in
the underlying models we present, where the dynamics is more complete, a
stronger degree of fine tuning is probably required both because gauge
couplings do not run linearly with momentum scale and because several
gauge groups become strongly coupled 
at essentially the same scale. Although the
reader may still find these tunings uncomfortable they are clearly much
less severe than those of the SM. We are agnostic and simply
wish to explore this paradigm of model building. 
Thankfully in these matters we will eventually be
instructed by experiment. 
One of the successes of
the flavour universal models is that they allow direct condensation of
all the electroweak doublets and hence masses for all the SM 
fermions without increasing the fine tuning. In the original top
condensation models, to generate the electron mass by a direct condensate
would have involved fine tuning of order $m_e/\Lambda$;
the suppression of the mass in the present models is the result of
the smallness of 
mass terms in a singlet sector. The structure also does not require very
large singlet masses to see-saw the electron mass small which again
would have been a source of fine tuning since it would have driven the
upper cut-off higher. 

\section{The Flavour Universal See-Saw Mechanism}
\label{sec2}

Our flavour universal see-saw model can be thought of as 
an extension of the top condensate see-saw model of Dobrescu and Hill 
\cite{bh}, 
which we briefly review next. 
That model, in addition to the top doublet, $Q_L = (t_L, b_L)$ and the $t_R$
contains the electroweak singlets $\chi_L, ~ \chi_R$ which have the same QCD 
and $U(1)_Y$ charges as the $t_R$. The EWS breaking is driven 
by the four fermion interaction
\begin{equation}
G \bar{Q}_L \chi_R \bar{\chi}_R Q_L,
\label{fourf}
\end{equation}
which, if the coupling $G$ is above critical,
generates an EWS breaking mass between $Q_L$ and $\chi_R$. 
This can be seen in the large $N$ approximation to the gap equation
\begin{equation}
{1 \over G} = {N \over 4 \pi^2} \left[ \Lambda^2 - m^2 \ln \left(
{\Lambda^2 \over m^2} \right) \right], 
\end{equation}
where $\Lambda$ is the scale above which the dynamics generating Eq.~(\ref{fourf}) 
resides, 
and $m$ is the dynamically generated mass. 
In addition all EWS invariant masses are allowed, that is $\bar{t}_R \chi_L$
and $\bar{\chi}_R \chi_L$. The resulting mass matrix below the 
EWS breaking scale
is then 
\begin{equation}
(\bar{t}_L, \bar{\chi}_L) \left( \begin{array}{cc} 0 
& m_{t_L \chi} \\ m_{t_R\chi} 
& m_{\chi \chi}
\end{array} \right) \left( \begin{array}{c} t_R \\ \chi_R \end{array}
\right)
\end{equation}
The value of $m_{t_L \chi}$ needed to provide all the EWS breaking 
vacuum expectation value
(VEV) may
be estimated using the Pagels-Stokar formula \cite{pagel}
\begin{equation}
v^2 \simeq {N_c \over 4 \pi^2} m_{t_L \chi}^2 \ln\left(\frac{\Lambda}
{m_{\chi\chi}}\right).  
\label{ps}
\end{equation}
Thus, for instance for $\Lambda/m_{\chi\chi}\simeq O(10)$,  
$m_{t_L \chi}\approx 600$~GeV. 
By appropriate choices of the
remaining two masses (e.g. $m_{t_R\chi} \simeq 1~$TeV and $m_{\chi \chi} \simeq 
3$~TeV) the lightest mass eigenstate of the mass matrix comes out as $175$~GeV, 
whereas
the most massive is $3~$TeV. In fact by suitable choices of the two singlet
masses the value of the lightest eigenstate mass and the electroweak breaking 
mass may be maintained even in the limit of decoupling the $\chi$ fermion by
taking $m_{\chi \chi} \rightarrow \infty$. Note that the electroweak breaking
mass in that limit is only for the left handed top doublet so does not violate
custodial isospin.

\begin{figure}
\center
\hspace*{0.7cm}
\psfig{figure=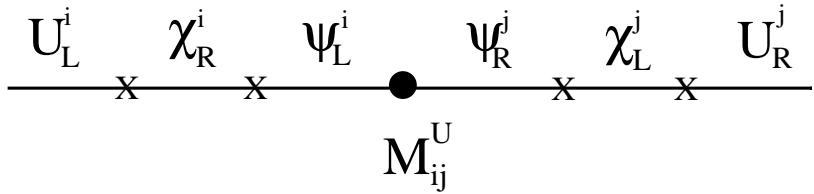,height=1.0in}
\caption{\small\em  The mixing that produces the up quark sector masses.
}
\label{fig1}
\end{figure}

There is a crucial difference between the top see-saw model and previous
top condensate models. 
That is that
the EWS breaking VEV of the left handed top is with a EWS singlet not the
right handed top. The right handed top also has a mass with a singlet
sector fermion. The SM top mass results from the mass mixing
in the singlet sector indirectly 
connecting the left and right handed tops. The
size of the top mass is no longer a direct consequence of its role
in EWS breaking but simply of some singlet mass structure. The top quark
is therefore in this sort of model no longer the essential fermion to be
involved in EWS breaking. Any SM fermion could play the
same role! In the see-saw of Eq. (3) the light mass eigenstate is
essentially given by $m_{t_L \chi} m_{t_R \chi} / m_{\chi \chi}$. 
Using the same see-saw but generating e.g. the electron mass would require
a very large value of $m_{\chi \chi}$. The upper cut-off on the theory
would have to be raised and with it the degree of fine tuning required
to generate the weak scale. To avoid this we will enlarge the singlet
sector. Consider for example the generation of the top quark mass. 
We introduce 
the additional electroweak singlet fields $\chi_L$, $\chi_R$ but also
$\psi_L$, $\psi_R$. We assume these extra singlets are all coloured under
the usual SU(3) QCD group and have the same hypercharge as the $t_R$.
The model then induces four fermion interactions of the form
(suppressing colour 
indices) 
\begin{equation}
\bar{Q}_L \chi_R \bar{\chi}_R Q_L
\label{top4f}
\end{equation}
In analogy with Eq.~(\ref{fourf}), these interactions become critical and
break EWS.  
We also allow the singlet mass terms $\bar{t}_R \chi_L$
and similarly $\bar{\chi}_L \psi_R$ and $\bar{\chi}_R \psi_L$. 
Finally we 
include the mass term $M^U \bar{\psi}_L \psi_R$, which provides 
the only connection between the left and right handed top. Note that we
have not included all possible singlet masses compatible with the gauge
symmetries; effectively we have included one massive fermion
($\bar{\chi}_L \psi_R$) that couples to the $t_L$ and another massive
fermion ($\bar{\chi}_R \psi_L$) that couples to the $t_R$. The two
massive fermions then have small mass mixings from the  $\bar{\psi}_L
\psi_R$ masses which are the only couplings between these
sectors. Although this may seem somewhat ad hoc at this stage the models
of Sections 2 and 3 naturally produce this mass structure. The important
point here is that there is some singlet mass structure that can
reproduce the SM fermion masses.
The mass matrix  takes 
the form 
\begin{equation}
(\bar{Q}_L, \bar{\chi}_L, \bar{\psi}_L) 
\left( \begin{array}{ccc} 0 & m_1 & 0\\
m_2 & 0 & m_3  \\  0 & m_4 
 & M^U\end{array} \right)
\left( \begin{array}{c} t_R \\ \chi_R\\ \psi_R \end{array} \right)
\end{equation}
where $m_1$ is the EWS breaking mass. 
In general the matrix is complicated but the pattern of masses resulting 
can be seen by taking the decoupling limit. We imagine that the largest masses 
are $m_3$ and  $m_4$ which bind the $\psi$ and $\chi$ fermions into two 
heavy Dirac fermions.
We may then treat the remaining masses as perturbations. The SM 
top  mass results from the diagram in Fig.~\ref{fig1} 
and the mass is proportional to the singlet mass $M^U$. 

This singlet sector mass structure is more readily convertible to
include the other SM fermions. We will write the SM
quarks as $Q^i_L$, $U_R^i$ and $D_R^i$ where  $i$
is a generation index. We may endow the fields $\chi$ and $\psi$ with a
generation index and the
mass matrices $m_1-m_4$ and $M^U$ with structure in the generation
space.  A four fermion operator of the form of Eq.~(\ref{top4f}) for
each left handed doublet will break EWS through a condensate for each
flavour. We shall assume that these operators give rise to flavour
universal VEVs since we expect the four fermion operators to arise from
some flavour universal broken gauge interaction. The precise way in
which this dynamics will be implemented depends on the choice of broken
gauge symmetry and we will discuss two possibilities: a gauged {\em family}
symmetry and the gauging of the complete {\em flavour} symmetry.
The dynamical implementation of these models and their
consequences are discussed in detail in the next two sections.

The important point of the present  structure is that the masses of the 
light eigenstates
are proportional to the singlet mass matrix $M^U_{ij}$. Thus, the 
suppression of  the up quark mass relative to the top quark mass only
requires us to have a relatively smaller mass element in the $\psi$ mass
matrix. The heaviest fermion masses ($m_3,m_4$) do not contribute to the
generation of the structure in the SM masses 
and hence the upper cut-off of the
theory can be left  unchanged even when we incorporate the lightest fermion
masses. There is therefore no increase in the degree of fine tuning in
the model if the up quark participates in EWS breaking.
At least at this stage in the model building, 
the three 
generations of singlet fermions $\psi$ reflect the same mass hierarchy
problem present in the spectrum of SM fermions.
The important point is that the origin of the symmetry breaking mass splittings
have been deferred to this EWS singlet sector. 
It is not our intention 
to address their origin here, though one could imagine 
generating these masses dynamically, via additional interactions 
residing at even higher energies, such as  
some singlet sector extended 
techincolour model or tumbling gauge structure.

Extending this scenario of mass generation to the isospin $-1/2$ quark sector 
is straightforward. 
We introduce the electroweak singlets $\omega_L^i$, $\omega_R^i$, as well as 
$\xi_L^i$, $\xi_R^i$, all with the same hypercharge assignments as 
right-handed down
quarks. A four fermion interaction $\bar{Q}_L^i \omega^i_R
\bar{\omega}_R^j Q^j_L$ will dynamically generate the condensate
$\langle \bar{Q}_L \omega_R \rangle$ contributing to EWS breaking. 
The singlets $\xi$ have a mass matrix $M_{ij}^D$ which gives the 
connection between the left and right handed down-quark sectors. 
$M_{ij}^D$ generates the pattern of 
down quark masses through diagrams analogous to the one in Fig.~\ref{fig1}. 

In this naive model with a four fermion interaction and explicit singlet
masses, a CKM matrix may be easily incorporated. There are in total seven
chiral family symmetries of the singlet fermions broken by six singlet mass
matrices. These matrices can not therefore in general be diagonalized
simultaneously and the model contains mixing angles that will feed into
the SM sector.   

The electroweak singlets, $\chi$ and $\psi$ in  the up sector, and 
$\omega$ and $\xi$ in the down sector,  have large
masses relative to the weak scale (of order a few TeV)
and may be considered to be decoupled.
The dynamical breaking of electroweak symmetry by the condensates between 
the SM left-handed fermions and the $\chi_R$, $\omega_R$ singlet fermions 
will generate a scalar sector of the low energy theory \cite{topcond}. 
The precise form
of the scalar sector is model dependent. We will discuss two examples of
this sector in the following two sections where the origin of the
dynamics is more concretely specified. 
However, it is possible to make a general point about the mass scale governing the
masses of the scalar sector in relation to that of the top see-saw
model.
The important difference is that the EWS breaking VEV is flavour universal. 
In order to see this difference
we notice that the Pagels-Stokar formula is  given by
\begin{equation}
v^2=\frac{N_f}{4\pi^2}\; \;m_1^2\; \ln \left({\frac{\Lambda}{m_3}}\right)~,
\label{ps2}
\end{equation}
where 
$N_f$ denotes the number of fermion flavours forming a condensate. 
For the top-condensate model, $N_f=N_c$ and we recover Eq.~(\ref{ps}). 
On the other hand, for example in a model where all the quarks
participate equally in EWS breaking 
$N_f=6N_c$. 
For instance, for $\ln{\Lambda/m_3}\simeq{\cal O}(1)$ one 
obtains $m_1\simeq 360~$GeV, 
whereas larger values of the ratio 
$\Lambda/m_3$ result in $m_1\simeq 250~$GeV or even lighter.   
This EWS breaking mass is approximately 
three times smaller than in the top see-saw model. 
If one naively computes the Higgs boson mass as
in the large $N$ approximation to the four fermion theory 
the Higgs mass may be extracted from a bubble resummation and is found to be 
\begin{equation}
m_h \simeq 2\;m_{1}~,
\label{mhnjl}
\end{equation}
implying  Higgs masses in the range $m_h\simeq (500-700)$~GeV. 

When the EWS breaking is driven by four fermion interactions it is easy
to include the leptons ($L^i_L$, $N_R^i$
and $E_R^i$) in the same pattern as the quarks. Additional EWS
singlets are introduced analogous to $\chi^i$, $\psi^i$, $\omega^i$ and
$\xi^i$ but with the SM interactions of the right handed
leptons. Four fermion interactions analogous to those of (5) drive
condensates involving $L^i_L$ and the massive singlet sector. The standard
model lepton masses come again from the mass matrices linking the
singlet $\psi$ fields. 
If the  leptons participate equally in the EWS breaking condensates, then
$N_f=6N_c + 6$ in the Pagels-Stokar formula, and the 
Higgs masses are controlled by a even lighter scale ($m_h \simeq 400-650GeV$). 

One of the benefits of this class of  models is that there are only very small
deviations from the SM values of electroweak precision
variables such as the $S$ and $T$ parameters. The $S$ parameter
essentially counts the number of electroweak {\em doublets}. The flavour
universal model has only extra electroweak {\em singlet} fields so makes no
extra contribution. By taking the singlet mass terms $(m_3, m_4)$ very large
one may decouple them from the low energy dynamics (perhaps at the expense of
increased fine tuning to generate $v$) or equivalently we may say that
the physical heavy fermions have only a very small admixture of the
electroweak doublets in them. In this limit the isospin breaking of the
heavy fermions decouple from the $T$ parameter \cite{bh}
leaving just the
contribution from the light SM fermions with the standard
masses. As this limit is relaxed all the  mass eigenstates, the heavy and
the light, will contribute to $T$ with by far the largest contribution
from the top quark and its singlet partners. The calculation of $T$ is very
similar to that of the top see-saw model and the additional
contributions to $T$ are easily controlled to be of the order the
experimental bounds for values of masses for the heaviest fermions of 
order $(1-5)~$TeV.

In the next two sections we present specific examples of the flavour universal 
see-saw mechanism, where the broken gauge dynamics is
made explicit. In one case it results from a broken gauged {\em family}
symmetry, in the other from a larger 
broken gauged {\em flavour} symmetry.
A crucial aspect of theories with gauged flavour
symmetries is to avoid the tight constraints on flavour changing neutral
currents (FCNCs). We will provide  models that 
achieve this by maintaining a GIM mechanism above
the electroweak scale and provide existence proofs that such gauge
dynamics is possible.
We also discuss the phenomenology of these models, 
considering  the consequences of the broken flavour symmetry as well as 
the scalar and pseudo-scalar sector of the models.

\section{Dynamics From Broken Family Symmetry}
\label{sec3}

Our first explicit example of a flavour universal EWS breaking model
will assume that the family symmetry of the left handed 
SM fermions is gauged and broken at a scale of a few
TeV. For simplicity we will restrict ourselves to the quark sector. The
$\chi_R^i$ and $\omega_R^i$ fields are assumed to also transform under
this gauge group. The
broken gauged family interactions result in the four fermion
interactions
\begin{equation}
\bar{Q}^i_L \chi_R^i \bar{\chi}_R^j Q^j_L + \bar{Q}^i_L \omega_R^i 
\bar{\omega}_R^j Q^j_L
\label{fam4f}
\end{equation}
where the family index is summed over. These interactions will generate
the EWS breaking VEVs between the left handed quark multiplets and the
singlet fermions. 

The singlet sector masses are assumed to be in place to generate the
quark mass matrices discussed in Section~2.

The scalar sector of the theory is expected to be large. The quark
condensates are breaking an $SU(6)_L \times SU(6)_R$ chiral symmetry of
the quarks and $\chi_R, \omega_R$ fields to the vector subgroup. To
represent this as a Higgs model we must have a Higgs field transforming
as a ($6, \bar{6}$) under the flavour group. There are thus 72 real
scalars. As a result of the chiral symmetry breaking 
there are $35$ NGB of which three are eaten to give masses to the 
$W$ and $Z$ gauge 
bosons. The remaining $32$ pseudo-scalars are naively  
massless but acquire masses through SM 
gauge interactions. They correspond to a color-octet, $SU(2)_L$ singlet and a
color-octet $SU(2)_L$ triplet.
Their masses due to the strong interactions can be 
computed to be \cite{pngbmass}
\begin{equation}
m^2_\pi = 3\alpha_s\; M^2~, 
\label{mpi}
\end{equation}
where $M$ is a high energy scale.  For instance, for $M\simeq O(1)$~TeV, we obtain
$m_\pi\simeq (500-600)~$GeV.  They couple to SM quarks with couplings 
proportional 
to $m_q/f_\pi$ ($f_\pi$ is the pNGB decay constant $\sim v$). 

As discussed in Section~2 the non-NGB scalars are expected to have 
masses in the $(500-700)~$GeV range. One of these, a pseudo-scalar,
is the would be NGB of
the $U(1)_A$ symmetry, the remaining 36 scalars break down into the same
SM gauge multiplets as the NGBs. 
We leave a more precise computation of
the spectrum analogous to that of \cite{topcond} 
for future investigation since it will be involved. 

At the level of Eq.~(\ref{fam4f}) the leptons may also be included in the
interaction that breaks EWS. If the leptons 
participate in EWS breaking, then the flavour symmetry
$SU(8) \times SU(8)$ is broken to the vector subgroup and there are 63
pNGB of which again only 3 are eaten. The additional 28 pNGB-s consist
of: two electroweak
triplets and two singlet leptoquarks with masses
\begin{equation}
m^2_\pi = {4 \over 3} \alpha_s\; M^2~, 
\label{mpi2}
\end{equation}
that is $m_\pi \simeq (300-400)$GeV; there is also a triplet and a
singlet of colour neutral pNGB. These latter pNGB do not receive masses from
the SM gauge interactions at one loop and are potentially in
conflict with experiment. This may be an indication that the leptons do
not participate in EWS breaking but instead receive their mass by some
radiative mechanism from the quark sector. In addition in this case 
there would be 65
massive scalars.

\subsection{A Dynamical Model of the Universal See-saw}

\begin{figure}
\center
\hspace*{0.7cm}
\psfig{figure=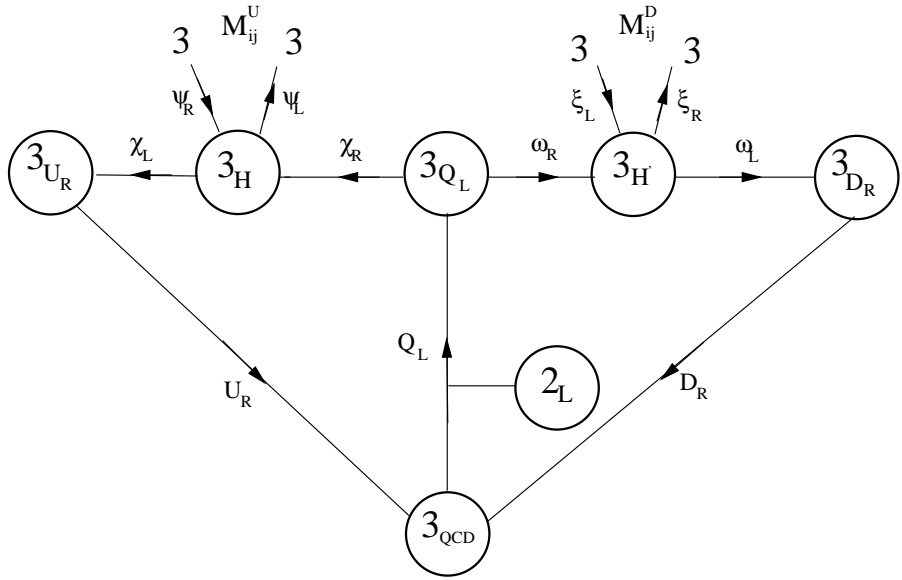,height=3.0in}
\caption{\small\em A dynamical model of the universal see-saw for the quark
sector displayed in moose notation. 
}
\label{fig2}
\end{figure}

In order to provide a more explicit, and renormalizable, 
model of the flavour universal
see-saw mass pattern described in Section~\ref{sec2}
we revive a model of top condensation
by King (in turn based on the technicolour model of Georgi
\cite{georgi}). 
The model is shown in
``moose'' notation in Fig.~\ref{fig2}. Recall that in moose notation a gauge group is 
represented by a circled number ($N$ for $SU(N)$) and left-handed 
fermions transforming in the (anti-) fundamental representation of that group
as lines with (out-) inward pointing arrows. Global symmetry groups are 
un-circled numbers. For simplicity we only consider the quark sector of the
theory until the end of  this section.

In Ref.~\cite{king} the dynamics was assumed to occur in the following fashion.
We concentrate on the isospin $+1/2$ sector for ease of discussion. 
The $SU(3)_H$ ``hypercolour'' 
gauge group under which $\chi$ and $\psi$ transform were assumed to
get strong generating the condensates $\langle \bar{\chi}_L 
\psi_R \rangle$ and
$\langle \bar{\chi}_R \psi_L \rangle$ 
and breaking the two chiral  family symmetry
$SU(3)$ groups to global groups. The formation of these condensates is not
the preferred breaking pattern when the family symmetry groups are weakly
coupled. 
However, there is some evidence that the effective potential favours
this pattern when the family symmetry groups are strongly coupled 
at the symmetry breaking scale. 
This hypothesis was first put forward by Georgi~\cite{georgi}, 
elaborated on in Ref.~\cite{evans}
and disputed in Ref.~\cite{luty}. 
We shall assume that this pattern of condensation is correct.
In the model of King these condensates were the only ones to form at this scale
and served the purpose of breaking the family gauge groups leaving strongly 
coupled ``flavourons''. The flavourons of the two chiral family groups
mixed through the $\bar{\psi}_L \psi_R$ mass matrix and the result was
four fermion interactions between the isospin $+1/2$ quark sector with the strongest
channel proportional to the largest eigenvalue of the $\psi$ mass matrix. This
strongest interaction was assumed to generate a top quark condensate 
$\langle \bar{t}_L t_R \rangle$. 

We now propose an alternative possibility. Since the family symmetry groups
are strongly interacting and close to their chiral symmetry breaking
scale when the 
hypercolour group breaks its chiral symmetries, it is not clear that the
$\chi$ and $\psi$ fields should be integrated out of the dynamics of
the family gauge groups. It is very plausible that the condensates
$\langle \bar{U}_R^i \chi_L \rangle$ and $\langle \bar{U}_L^i \chi_R^i 
\rangle$ form as well. If we assume this, then the strong dynamics
immediately produces the pattern of masses needed for the flavour universal
see-saw model. Precisely how large the three different condensates are
is a sensitive matter of the three gauge groups interrelated strong 
interactions. We shall assume that by appropriate choices of couplings at 
the hypercolour group's strong scale the desired values can be obtained.
There is of course, as in all top condensate models envisaged to date, a
serious problem of fine tuning in the assumption that many gauge groups
become strongly interacting at almost an identical scale. We offer no 
excuses (though it might be more natural if the gauge dynamics are walking 
in nature) but put forwards this model as a simple existence proof of 
dynamics that might give rise to the required effective theory.

The existence of these additional condensates also breaks the $SU(3)$ colour
group and the hypercolour group to their vector subgroup that then plays the 
role of the low energy QCD group. There are thus flavour universal colourons
analogous to those in the model of Ref.~\cite{colorons}. The NGB associated
with the condensate $\langle \bar{U}_R^i \chi_L^i \rangle$ are eaten
in the breaking of these two groups to the vector subgroup. The
NGB from the condensate  $\langle \bar{U}_L^i (\chi_R^i+\omega^i_R) 
\rangle$ are those already discussed above. Finally there are the 36 NGB
associated with the $\langle \chi \psi \rangle$ condensates of which 16
are eaten to give masses to the two sets of family symmetry group gauge
bosons. The remaining NGBs will acquire masses from the explicit mass
matrix $M^U_{ij}$ and the family gauge interactions that explicitly break the
chiral symmetry group. However, the mass matrix is off diagonal in the
basis of the condensate and the gauge interactions chiral, so these
symmetry breakings will only contribute to the pNGB 
masses at second order. These EWS
singlet pNGBs may therefore only have masses of the order of a few GeV. 
We will live with their existence since they only occur in 
the singlet sector of
this model, which is in any case only intended 
to illustrate the concept of a flavour universal seesaw. We do not believe 
they are necessarily a robust signature of this class of models.

The bottom quark sector of the model behaves similarly to the top quark 
sector and the relative mass differences in the final spectrum are
trivially the result of different mass matrices included by hand for 
the $\psi$ and $\xi$ singlet fermions of the two sectors. 
Including mixing angles is non-trivial \cite{georgi}. 
They will result if the singlet fermion mass matrices in the up and 
down sectors, $M_{ij}^U$ and $M_{ij}^D$  are not simultaneously diagonal. 
As shown in Fig.~\ref{fig2}  the model
has sufficient global symmetry to simultaneously diagonalize both mass 
matrices and there will be no mixing angles. An attempt to remedy this 
problem by including yet higher energy scale dynamics can be found in 
Ref.~\cite{georgi}. We wish to defer questions of such high scale dynamics
and simply note that there does not appear to be any problem, in principle, 
preventing the generation of  mixing angles from the mass matrices  of 
the singlet sector. 

The model has flavour symmetries gauged at a relatively low scale, ${\cal
O}(1)~$TeV,   which might naively be expected to generate FCNCs.
However, as promoted in Ref.~\cite{georgi}, the model
has the same chiral flavour symmetries as the SM 
even above the weak
scale (that is $[SU(3)\times
U(1)]^3$) and hence a GIM 
mechanism that allows the neutral currents to be written in diagonal
form. FCNC effects will be generated with the inclusion of the generic mass
matrices $M^{U/D}{ij}$ but the leading effects are through
the quark masses themselves, and therefore  simply correspond to the SM
contributions. The largest additional contributions are from mass
mixings of the left and right family gauge groups that lead to four fermion operators
between  the up (down)-type
quarks mixing through $M^D$ ($M^U$) which rotations on those fields cannot 
diagonalize. The leading such term contributing to $\Delta S = 2$
processes is for example
\beq
 {1 \over \Lambda_H^6}
\bar{Q}_L^i (M^U M^{U \dagger})_{ij} Q^j_L  \bar{Q}_L^k (M^U
 M^{U \dagger})_{kl}  Q_L^l
\eeq
In addition to the suppression by the gauge boson mass these
contributions are suppressed by a factor of the SM mass
mixing matrix element over 50 GeV to the fourth power. They are therefore much
smaller than the SM contributions. 

The inclusion of leptons in the model is not so straightforward. 
Naively one would
simply repeat the model of the quark sector with the SM leptons
transforming under the same family symmetry groups. 
However, this would  
introduce unacceptably large lepton 
flavour violating effects (e.g. $K^0 \rightarrow e \mu$)
since two multiplets would now transform under
a single family symmetry group only broken at the few TeV scale. 
As an alternative, one could simply replicate the whole moose model for the lepton sector.
Additional $\chi$
and $\omega$ fields transforming under their own hypercolour groups would
need to be added and the lepton mass matrices would come out proportional
to the mass matrix put in between the new $\omega$ fields. The only problem is
that to cancel anomalies the hypercolour group must be a U(1) group
which is not asymptotically free. Rather than make the model more
baroque to force the leptons in, we will leave the model at this point
since the similar model of the next section succeeds more naturally 
in including the leptons,  
although not without similar problems.

\subsection{Phenomenological Considerations}
\label{pheno}
In this section we discuss the  phenomenological constraints on the 
family symmetry generated 
flavour universal see-saw model as well as its  potential signatures at future
experiments. Here we concentrate on the model giving quark masses and EWS breaking, 
leaving the 
discussion of lepton masses for the next section. 
The model described in the previous section contains several new states. There are 
the fermion singlets,  $\chi_{L,R}$ and $\psi_{L,R}$ connected to the up-quark sector, 
and $\omega_{L,R}$ and $\xi_{L,R}$ to the down-quark sector. As mentioned 
in Section~\ref{sec3}, 
these acquire large masses as a result of the condensates formed among them, 
induced by the 
strong up and down hypercolour groups. This, together with the fact that they 
are electroweak 
singlets, implies that they are not relevant to the low energy phenomenology. 
They are, at 
this stage, simply conduits to communicate between the left and right-handed 
quark sectors
and to mix with the left-handed quark doublets and generate EWS breaking 
at the right scale.

\subsubsection{Flavourons}

The breaking of the three gauged family groups $SU(3)_L$, $SU(3)_{U_R}$ and 
$SU(3)_{D_R}$ 
by the formation of the necessary singlet fermion condensates, 
implies the existence of three new sets of massive gauge bosons. These are 
family octets 
coupling separately to left-handed quark doublets, right-handed up 
and right-handed down quarks. 
The masses of these ``flavourons''
are expected to be similar and are determined by the scale where the hypercolour
groups break the family symmetries. 

The couplings of the flavourons to the SM fermions are given by 
\begin{equation}
{\cal L}_f = -g_L \;L_\mu\; \bar{Q_L^i}\gamma^\mu t^A_{ij} Q_L^j
-g_U \;R^U_\mu\; \bar{U_R^i}\gamma^\mu r^A_{ij} U_R^j
-g_D \;R^D_\mu\; \bar{D_R^i}\gamma^\mu s^A_{ij} D_R^j~~,
\label{flacou}
\end{equation}
where $L_\mu$, $R^U_\mu$ and $R^D_\mu$ are the 
flavourons corresponding to the gauge 
family groups $SU(3)_L$, $SU(3)_{U_R}$ and $SU(3)_{D_R}$ respectively; 
and $t^A$, $r^A$ 
and $s^A$ are their respective generators, with $A=1,..,8$. They simply are 
$t^A_{ij}=r^A_{ij}=s^A_{ij}=\lambda^A_{ij}/2$, with $\lambda^A$ 
the generators of SU(3).
The fact that flavourons are flavour-octet gauge bosons seems to indicate 
the existence of non-diagonal vertices. 
However these do not result in FCNC processes
due to the 
presence of the remnant
global $SU(3)$ family symmetries after the breaking of the gauged ones. 
These approximate global symmetries allow us to implement the GIM 
mechanism along the lines of the models of Ref.~\cite{georgi} and 
are only broken by fermion mass terms.

The flavourons acquire masses at the scale $\Lambda_f$ where the 
hypercolour groups break the family symmetry. Thus, they are expected 
to be in the TeV 
range. On the other hand, their couplings to fermions must be strong enough
to generate the $\langle \bar Q_L\chi_R\rangle$ and $\langle \bar Q_L\omega_R\rangle$
condensates that break the EWS, as well as 
the $\langle \bar U_R\chi_L\rangle$
and $\langle \bar D_R\omega_L\rangle$ necessary to obtain
the SM quark masses. We compute the criticality 
condition in the NJL approximation.  
Defining $\kappa_a\equiv g_a^2/4\pi$, with $a=L,U,R$, the couplings must satisfy
\begin{equation}
\kappa_a \geq 2\pi\left(\frac{N_g}{N_g^2-1}\right)~,
\label{crit}  
\end{equation}
where $N_g$ refers to the number of generations.
The condition Eq.~(\ref{crit}) has important phenomenological consequences. 
For instance, it 
gives a lower bound for the production cross section, for a fixed flavouron mass.
At the same
time it implies that the flavourons' widths are rather large. 
For instance the flavourons coupled to the left-handed quarks have a width given by
\begin{equation}
\Gamma_f \simeq ~\frac{\kappa_L}{6}M_F, 
\label{width}
\end{equation}
which implies that the minimum width, given by  
the critical value of $\kappa_L$, is approximately $40\%$ of its mass 
making it difficult for them to be detected as clear mass bumps  in  
$p\bar p$ collisions at the Tevatron.
The situation is somewhat better for the flavourons coupled to the right-handed 
up and down quarks. Their minimum widths are about half that of 
Eq.~(\ref{width}).  
On the other hand and for all cases, the large couplings in Eq.~(\ref{crit}) 
ensure large excesses in hadronic production of all flavours. 
The fact that flavourons
are chirally coupled 
may produce a very  distinct signal in high transverse momentum jets. 
At energies below the flavouron mass these excesses mimic the behavior of contact 
terms with specific chirality. The limits from the Tevatron Run~I data as well 
as the reach of Run~II is currently under study. 

The presence of strongly coupled gauge bosons 
may be constrained by their contributions to the $T$ parameter. 
As pointed out in Ref.~\cite{cdt}, this is so even when the interaction itself is
isospin conserving. The exchange of the new gauge bosons gives a two loop
radiative correction to the isospin violating 
contributions to the $\rho$ parameter coming from  one-loop diagrams 
involving the SM fermions.  
We first notice that these types of contributions to $T$ are induced only by the 
gauge bosons that couple to left-handed fermions. Thus, only one of the three 
flavourons must be considered.  
The relevant contribution to the $T$ parameter is given by 
\begin{equation}
T=\frac{4\pi}{s^2\theta_Wc^2\theta_W M_Z^2}\left(
\frac{\Pi_{LL}(m_t,m_b)}{2}-\frac{\Pi_{LL}(m_t,m_t)}{4}
\right)~,
\label{tdef}
\end{equation}
where $\Pi_{LL}(m_t,m_b)$ and $\Pi_{LL}(m_t,m_b)$ refer  to the left-handed
vacuum polarizations involving one top and one bottom quark , and two top 
quarks in the loop respectively, and they are evaluated at zero momentum 
transfer. 
Following \cite{cdt} we will approximate
the calculation of the two-loop diagram by a product of two one loop diagrams
obtained after shrinking the flavouron propagator. Then, the vacuum polarizations
are
\begin{equation}
\Pi(m_t,m_b)=\frac{\Pi(m_t,m_t)}{4}\simeq -\frac{1}{16\pi^3}m_t^4\left(
\log\frac{\Lambda_f}{m_t}\right)^2\frac{\kappa_L}{M_F^2}~,
\label{vacpol}
\end{equation}
where the vacuum polarizations were approximated by their divergent 
behavior given by the leading logarithm. 
This results in 
\begin{equation}
T=\frac{m_t^4}{s^2\theta_Wc^2\theta_W M_Z^2}\,\frac{1}{8\pi^2}
\left(
\log\frac{\Lambda_f}{m_t}\right)^2\frac{\kappa_L}{M_F^2}~.
\label{tres}
\end{equation}
This constraint is compatible with $\Lambda_f$ being of the 
order of the TeV scale. For instance, if we take $\log(\Lambda_f/m_3)\simeq 1$, 
and we assume $m_3\simeq 1~$TeV, then the induced $T$ parameter is 
\begin{equation}
T\simeq 0.06\frac{\kappa_L}{M_F^2}~.
\label{bound}
\end{equation}
The current determination of $T$ from electroweak precision measurements
gives~\cite{pdg} $T=(-0.11\pm 0.16)$, assuming a $300~$GeV Higgs mass. 
Thus, we see from Eq.~(\ref{crit}) and Eq.~(\ref{bound}) that flavouron masses 
in the $1~$TeV range and above are not in contradiction with the 
electroweak precision data.

\subsubsection{Colourons}

The flavourons induce four-fermion interactions 
between quarks and singlet fermions
which are supercritical and result in EWS breaking as well as in the 
breaking of the 
hypercolour groups in such a way that 
$SU(3)_{H}\times SU(3)_{H'}\times SU(3)_c$ 
breaks down to the ordinary QCD interactions. 
This breaking pattern leaves two sets
of color-octet massive gauge bosons, the ``colourons'', each of them 
coupling separately
to the up and down quarks. 

The two sets of colourons in this model present some similarities with the 
massive color-octet gauge boson in the model of Ref.~\cite{colorons}. 
As with the flavourons, they will give contributions to the
renormalization of SM $\rho$ parameter. The resulting bounds
are similar to those discussed for the flavourons and are in the TeV range
\footnote{The essential difference between the two cases 
resides in the fact that the same colourons couple to left and right handed
fermions. As a result, the effect in Eq.~(\ref{bound}) is larger by a factor 
of $3$ for colourons.} 
Moreover, and similarly to the case of the flavourons discussed above, the 
colouron couplings
must be supercritical in order for the hypercolour groups to generate
 mass terms in the 
singlet sector. 
This constraint, here once again, will imply large production cross sections
but also large widths. The width of either of the color octets is given by 
$\kappa_C~M_C$, with $\kappa_C$, the coupling of the corresponding colouron, 
reflecting
the embedding of $QCD$ in the hypercolour group and above the critical
value.   
The main difference with the flavouron case is that colourons interfere 
with the QCD 
gluon-mediated processes. 
The consequences of this interference at hadron colliders were first studied 
in Ref.~\cite{parkhill} in the context of Topcolor models, and later investigated in 
Ref.~\cite{colorons} for the case of flavour universal color-octet gauge bosons.  
The phenomenology of the colouron sector of the model is very similar to the one described
in this latter work.

\subsubsection{Scalars and pNGB}

Finally, in the model of Fig.~\ref{fig2}
the breaking of the $SU(6)_L\times SU(6)_R$ chiral symmetry under which 
$Q_L$, $\chi_R$ and $\omega_R$ transform, by a four fermion interaction 
implies the existence of a $(6, \bar{6})$ Higgs boson. The 72 real
degrees of freedom
split into 3 NGBs eaten by the $W$ and $Z$ bosons, 32 color -octet pNGB with
masses estimated above to be in the few hundred GeV range, 
one pNGB acquiring a large mass due to the $U(1)$ anomaly,  
and 36 scalars in the Higgs sector with masses estimated to lie between
$(500-700)~$ GeV.  

These states couple to the SM fermions through Yukawa
interactions of order $m_f/f_\pi$. 
In fact they are precisely the
lightest states found in a one family technicolour model without 
techni-leptons, and are subject to the similar low energy constraints. 
The most significant bounds on this scalar sector come from  
the measurements of $\Gamma(Z\to b\bar b)$
and the $b\to s\gamma$ branching ratio. The dominant
contributions correspond to the color-octet scalars and pseudo-scalars. 
For instance, the effect of considering one of these color-octets in the 
$b\to s\gamma$ transition, when combined with the $3\sigma$ interval from 
the most recent CLEO measurement of the inclusive rate~\cite{cleo1},
${\rm Br}(b\to s\gamma)=(3.15\pm0.35\pm0.32\pm0.26)~10^{-4}$,
translates  into a lower bound of $440~$GeV on the mass of the charged 
states. 
The effects of adding all the scalar and pseudo-scalar states may 
tighten the bounds, depending on the scalar masses. However, large 
cancellations 
are also 
possible among the various contributions and the SM 
$b\to s\gamma$ amplitude.  
The additional information from 
$b\to s\ell^+\ell^-$ decays, to be available in the near future from the next 
generation of $B$ physics experiments, will elucidate this possibility.
On the other hand, 
the $R_b$ bounds are not affected by the these cancellations.
For instance, the $3\sigma$ constraint on the color-octet states 
(assuming equal masses for scalars and pseudo-scalars) translates into 
a mass limit of  $ 700~$GeV.  However, if the central value of $R_b$ is taken to be 
the SM one, the mass constraint is considerably smaller, although still a few hundred
GeV.

\section{Dynamics From Broken Flavour Symmetry}
\label{sec4}

There is an alternative possible flavour structure for the four fermion
interactions driving EWS breaking that allow the leptons to condense and
hence acquire masses but
without giving rise to the light pNGB found in Sections~2 and~3 above. 
That is to allow the index $i$ on
$Q_L, U_R$ and $D_R$ to run over the full possible set of fermion flavours. 
In other words, in this model the index 
$i$ runs over the full $SU(12)$ flavour symmetry of the SM  
where the $12$ flavours are three families of three colours of
quarks and the three families of leptons. We introduce additional
$\chi^i$ ($\omega^i$) and $\psi^i$ ($\xi^i$) 
electroweak singlets with a similar pattern of
singlet masses to those of the model of Section~2. The colour and
hypercharge interactions are weakly gauged subgroups of the $SU(12)$
flavour symmetry of the SM particles and the electroweak
singlet sectors. The four fermion interaction 
\beq\bar{Q}^i_L \chi_R^i
\bar{\chi}_R^j Q^j_L + \bar{Q}^i_L \omega_R^i
\bar{\omega}_R^j Q^j_L
\eeq
drives electroweak symmetry breaking through the
condensate $\langle \bar{Q}^i_L (\chi_R^i+ \omega^i) \rangle$. 
The $SU(2) \times
SU(2)$ chiral symmetry of the fermion doublets is broken to the vector
subgroup producing three Goldstone bosons that are all eaten by the $W$ and
$Z$ bosons. In this case there are two charged Higgs bosons, 
two neutral Higgs and a massive pseudo-scalar, 
as in a standard two Higgs doublet model, with masses
of order $(400-650)~$GeV and no pNGBs,   
which makes this model phenomenologically more appealing 
than the one in the previous section. 

\begin{figure}
\center
\hspace*{0.7cm}
\psfig{figure=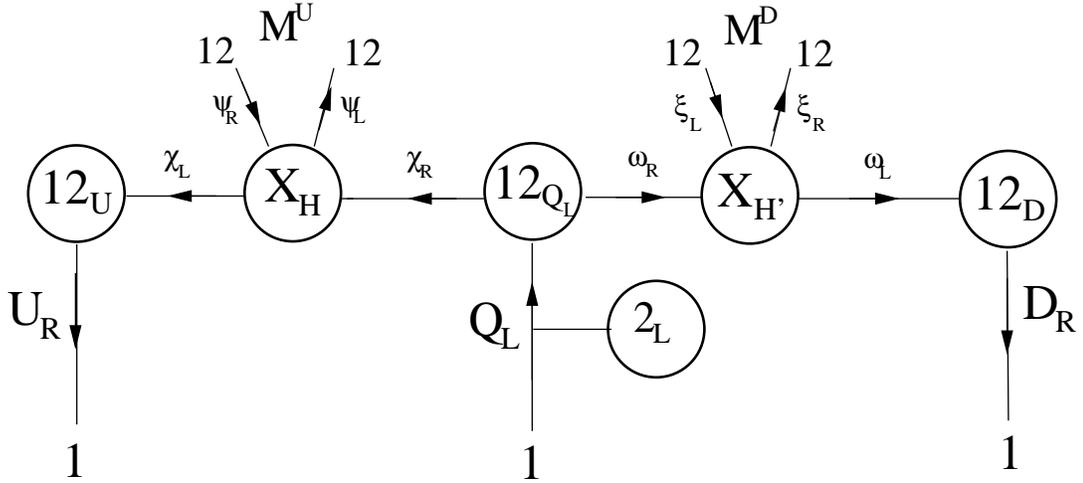,height=2.50in}
\caption{\small\em A flawed dynamical model of the universal see-saw resulting
  from broken flavour symmetry displayed in moose notation. 
}
\label{fig3}
\end{figure}

It is natural to promote this idea to a moose model based on that of
Section~3. One might propose the moose model of Fig.~\ref{fig3}, a variant of
the technicolour model of Ref.~\cite{randall}. 
We expect again that the 
dynamics is such that the $SU(X)$ hypercolour groups become strong and
break their chiral symmetries in the few TeV range. The chiral flavour
groups are also assumed to be strong at this scale. Amid these
dynamics the condensates $\langle Q_L^i (\chi_R^i + \omega_R^i) \rangle$,
$\langle U_R^i \chi^i_L \rangle$ and $\langle D_R^i \omega^i_L \rangle$ are
assumed to form. Mass matrices put in by hand for the $\psi$ and $\xi$
fermions provide the left right connecting structure necessary to
produce the SM fermion masses.  This is just the universal
see-saw mass scheme. 
The SM gauge interactions result as the vector subgroup of the
gauged SU(12) chiral flavour symmetries and any weakly gauged subgroup
of the $\psi$ and $\xi$ fermions global $SU(12)$ symmetry groups. We do
not show these weakly gauged subgroups in the figure for ease of
display. 
The electroweak symmetry breaking VEVs do not generate any pNGBs since
the $3$ NGBs arising from the breaking of the $SU(2)_L\times SU(2)_R$
chiral group are eaten by the $W$ and the $Z$ gauge bosons. 
However, the hypercolour dynamics will generate $575$ NGB, corresponding 
to the breaking of the $SU(24)_L\times SU(24)_R$ chiral symmetry of the 
$\chi$'s and the $\psi$'s.    
Of these, only $286$ 
are eaten by the broken flavour symmetry groups. The remainder are, just as
in the quark moose,
potentially light although they are singlets under the SM
interactions. 
As in that model, these light states are a result of the particular
implementation of the singlet sector and not an automatic consequence of the
flavour universal models.
The model again has a GIM mechanism that suppresses FCNCs.

A more serious problem with the present model resides in the fact  
that anomaly cancellation for the
gauged $SU(12)$ groups requires $X=1$. 
Attempts to make these groups
non-abelian by the inclusion of new EWS singlet fermions transforming
under the $SU(12)$ groups give rise to additional 
pNGBs, which may be unacceptably light 
(see Ref.~\cite{randall} for attempts in these directions). 

We present Fig~\ref{fig3}, although it is unsatisfactory, because it helps to
visualize the dynamics we envisage might be behind the model presented
first at the four fermion level. A flavour universal EWS breaking
scenario is possible without pNGBs and will most likely be associated
with a large, broken, chiral, gauged flavour symmetry of at least the
left handed SM fermions. The hypercolour
dynamics is not essential to the basic idea and may at least temporarily
be replaced by Higgs bosons in the absence of model building ingenuity.

\subsection{Phenomenological Considerations}

The broken flavour symmetry model is much cleaner at low energies than
the model of Section~3. There are no un-eaten NGB associated
with EWS breaking. The dynamics is expected to give rise to four scalar Higgs
bosons and a pseudo-scalar 
each with masses $(400-650)~$ GeV which  correspond to a standard two Higgs
doublet model with the ratio of Higgs VEV, $\tan \beta = 1$. 

The massive flavourons are similar to those of the previous model except
that they now also mediate interactions involving leptons. The 
GIM mechanism of the  model
suppresses FCNCs. 
One important modification with respect to the model of the previous section is
the criticality condition. The couplings $\kappa_{L,R}$ obey conditions similar to 
Eq.~(\ref{crit}) but with $N_g$ replaced by $N=12$. As a result the necessary  
couplings are considerably smaller than in the previous cases, which has  
important consequences in the phenomenology.   
For instance, the minimal flavouron contributions
to the T parameter in Eq.~(\ref{bound}) can now be smaller by roughly a factor of four, 
relaxing the mass bound by a factor of two.   
On the other hand, 
somewhat more stringent limits come from neutral current processes 
involving effective contact terms of the form $\ell\ell q\bar q$. 
We concentrate on $\ell=e$, where most of the experimental information is. 
The effective $eeqq$ interactions are   parametrized by the standard 
expression~\cite{eqpar}
\begin{eqnarray}
{\cal L}_{\rm NC}& = & \sum_{q}\left\{ \eta_{LL}^{eq}\;(\bar e_L\gamma_\mu e_L)
(\bar q_L\gamma^\mu q_L) 
+  \eta_{RR}^{eq}\;(\bar e_R\gamma_\mu e_R)(\bar q_R\gamma^\mu q_R) \right. \nonumber\\
 & &~~~\left.+\eta_{LR}^{eq}\;(\bar e_L\gamma_\mu e_L)(\bar q_R\gamma^\mu q_R)
+\eta_{RL}^{eq}\;(\bar e_R\gamma_\mu e_R)(\bar q_L\gamma^\mu q_L)\right\}~.
\label{eeqq}
\end{eqnarray}
The exchange of the $SU(12)$ flavourons generates the first two terms, whereas 
the $LR$ and $RL$ terms arise only as a consequence of one loop mixing  of the
$L$ and $R$ gauge bosons, which is suppressed by two powers of the masses, relative 
to the $LL$ and $RR$ coefficients. The $LL$ and $RR$ coefficients  take the form
\begin{eqnarray}
\eta_{LL}^{eq}&=&\frac{\pi\kappa_L}{M_{F_L}^2}~.\label{eta1} \\
\eta_{RR}^{ed}&=&\frac{\pi\kappa_R}{M_{F_R}^2}~. \label{eta2}
\end{eqnarray}
where $q=u,d$. The $SU(2)_L$ relations naturally resulting in this model imply that
the $RR$ interactions of the $u$ quark involve neutrinos instead of charged leptons.
This mismatch results in potentially dangerous contributions to which Atomic Parity
Violation (APV) as well as neutrino scattering experiments are especially sensitive.   
The most constraining bounds come from the APV experiments. The resulting
contribution to the atomic weak charge is given by~\cite{zepch}
\begin{equation}
\Delta C_q=\frac{v^2}{2}\left(\eta_{RR}^{eq}-\eta_{LL}^{eq}\right)~.
\label{dcq}
\end{equation}
Thus, assuming that the couplings and masses of the left and right handed 
sector are similar, the effect largely cancels in the $eedd$ interactions.  
This is not the case for $eeuu$. From the most recent measurements~\cite{zepch}
we obtain the $3\sigma$ bound 
\begin{equation}
\eta_{LL}^{eu}<0.22 
\label{etabound}
\end{equation}
which, for the critical value of the flavouron coupling translates into 
$M_{F}>2~$TeV. Looser bounds on $M_{F}$ are obtained from the 
$\nu N$ scattering experiments, as well as from HERA, LEPII and Drell-Yan processes
at the Tevatron. Thus, a scale of a few TeV for the dynamics of this model is not in 
contradiction with experimental observations.

\section{Conclusions}

We have presented model building ideas that extend the top condensate
see-saw model of Ref.~\cite{bh} to include realistic mass generation for all
the SM fermions. 
In these models 
all the electroweak doublets of the
SM play an equal role in breaking EWS, with their mass
differences resulting from different mass mixings in a heavy EWS singlet
fermion sector. 
This effectively separates the questions of EWS breaking and the origin
of fermion masses, deferring the latter to higher energy scales. 
As a result, the origin of the dynamics of fermion masses
is removed as a source of contamination of the precision 
electroweak variables $S$ and $T$. The
mass mixings of the EWS singlet fermions, that are the seed for the standard
model fermion masses, may be generated at much higher scales than the
weak scale without inducing further fine tuning in the production of the
weak scale. 

We presented two classes of models, based on broken gauged family 
symmetry in one case and broken gauged flavour symmetry in the other.
In both cases    
the resulting composite Higgs sector is relatively light, with masses in the 
range $m_h=(400-700)~$GeV. In gauged family models, the breaking of the large 
chiral symmetries in the SM fermion sector leads to the presence of 
a large number of scalars and pseudo-scalars with masses in this range, 
resulting in constraints from the measurements of $R_b$ and FCNC processes. 
On the other hand, the scalar sector of the broken flavour symmetry models
is more economical, with no pNGBs, and corresponds to the two Higgs doublet 
model with $\tan\beta=1$. 

Both scenarios involve extensions of the gauge sector and imply the existence of
massive gauge bosons associated with the breaking of either the family or flavour
symmetries, the flavourons. These are strongly  coupled to the SM fermions and 
thus have a very rich phenomenology at present and future high energy colliders. 
We have seen that the mass scale of the flavourons can be as low as a few TeV and still
satisfy all existing phenomenological bounds. 
Therefore, the experiments to take place at     
the Fermilab Tevatron Collider, as well as the CERN LHC 
will be sensitive to a variety of signals associated with these states, as well 
as with the colourons already present in other models of dynamical EWS breaking.

\section{Acknowledgments}

The authors are grateful to the Aspen Center for Physics where this work
was begun and to S. Chivukula, 
B. Dobrescu, C. Hill and J. Terning for useful discussions. This 
work was supported in part by the U.S.~Department of Energy 
under  
contracts $\#$DE-FG02-91ER40676 and $\#$DE-FG02-95ER40896 
and the University of 
Wisconsin Research Committee with funds granted by the Wisconsin 
Alumni Research Foundation.


\begin{thebibliography}{99}

\bibitem{TC} S. Weinberg, {\em Phys. Rev.} {\bf D19} 1277 (1979);
      L. Suskind, {\em Phys. Rev.} {\bf D20} 2619 (1979);
      E. Farhi and L. Susskind, {\em Phys. Report} 74 No.3 277 (1981);
      S. Dimopolous and L. Susskind, {\em Nucl. Phys.} {\bf B155} 237 (1979);
      E. Eitchen and K. Lane, {\em Phys. Lett.} {\bf B90} 125 (1980).

\bibitem{peskin} M.E. Peskin and T. Takeuchi,  {\em Phys. Rev. Lett.} 
         {\bf 65} 964 (1990);
         M.E. Peskin and T. Takeuchi,   {\em Phys.  Rev.} {\bf  D46} 381
         (1992).

\bibitem{topcond}  Y. Nambu, ``New Theories In Physics", Proc. XI Warsaw
Symposium on Elementary Particle Physics, (ed. Z. Adjuk {\it et al.}, publ.
World  Scientific, Singapore, 1989);
V.A. Miransky, M.Tanabashi and M. Yamawaki, {\em Phys. Lett.} {\bf
B221} (1989) 177; R.R. Mendel and V.A. Miransky, 
{\em Phys. Lett.} {\bf B 268} 384 (1991);
W.A. Bardeen, C.T. Hill and M.Lindner, {\em Phys. Rev.} {\bf D41}
1647 (1990).

\bibitem{tcatc}C. T. Hill, {\em Phys. Lett.~}{\bf B345}, 483 (1995). 

\bibitem{tcatc2}G. Buchalla, G. Burdman, C. T. Hill and D. Kominis, 
{\em Phys. Rev.~}{\bf D53}, 5186 (1996); G. Burdman and D. Kominis, 
{\em Phys. Lett.~}{\bf B403}, 101 (1997);
K. Lane and E. Eichten, 
{\em Phys. Lett.~}{\bf B352}, 382 (1995); K. Lane, {\em Phys. Rev.~}{\bf D54}, 
2204 (1996). 
 

\bibitem{cdt} S. Chivukula, B. Dobrescu and J. Terning, 
{\em Phys. Lett.~}{\bf 353}, 289 (1995); T.W. Appelquist,
N. Evans and S.B. Selipsky, {\em Phys. Lett.} {\bf B374} 145 (1996). 

\bibitem{bh} B. Dobrescu and C. T. Hill,{\em ~Phys. Rev. Lett.} 
         {\bf 81} 2634 (1998); 
         R.S. Chivukula, B.A. Dobrescu, H. Georgi and C.T. Hill, 
         {\bf hep-ph/9809470}. 

\bibitem{king} S. F. King, {\em Phys. Rev.~}{\bf D45}, 990 (1992).

\bibitem{georgi} H. Georgi, ``Technicolor and Families", Proc. 1990
International
         Workshop On Strong Coupling Gauge Theories And Beyond, (ed. T. Muta
         and K. Yamawaki, publ. World Scientific, Singapore,1991);
          H. Georgi, {\em Nucl. Phys.} {\bf B416} 699
         (1994).

\bibitem{evans}N. J. Evans, S. F. King and D. A. Ross, 
             {\em Z. Phys.~}{\bf C60}, 509 (1993).

\bibitem{randall} L. Randall, {\em Nucl. Phys.} {\bf B403} 122 (1993).

\bibitem{moose} R.S. Chivukula and H. Georgi, 
                   {\em Phys. Lett.} {\bf B99} 188 (1987);
                R.S. Chivukula, H. Georgi, L. Randall, 
                   {\em Nucl. Phys.} {\bf B292} 93 (1987). 

\bibitem{luty} M. Luty, {\em Phys. Lett.} {\bf B292} 113 (1992). 

\bibitem{colorons} R.S. Chivukula, A.G. Cohen and  E.H. Simmons, 
{\em Phys. Lett.~}{\bf B380}, 92 (1996); E. H. Simmons, {\em Phys. Rev.~}{\bf D55}, 
1678 (1997)
and M. B. Popovic and E. H. Simmons,  {\em Phys. Rev.~}{\bf D58}, 
095007 (1998).

\bibitem{pagel} H. Pagels and S. Stokar, {\em Phys. Rev.} 
{\bf D20} 2947 (1979).

\bibitem{pngbmass} M.E. Peskin, {\em Nucl. Phys.} {\bf B 175} 197 (1980).

\bibitem{pdg}J. Erler and P. Langacker, in ``Review of Particle Physics'', 
{\em Eur. Phys. J.}{\bf C3}, 1 (1998).

\bibitem{parkhill}C. T. Hill and S. J. Parke, {\em Phys. Rev.~}{\bf D49}, 
4454 (1994).

\bibitem{cleo1}Petr Gaidarev, for the CLEO collaboration, presented at the 
{\em ``Workshop on Heavy Quarks at Fixed Target, HQ98, Fermilab, October 9-12, 1998''}. 

\bibitem{eqpar}E. Eichten, K. Lane and M. Peskin, {\em Phys. Rev. Lett.~}{\bf 50}, 
811 (1982).
 
\bibitem{zepch} For a recent update see 
D. Zeppenfeld and K. Cheung, {\bf hep-ph/9810277} 
in {\em ``Proceedings of the WEIN Symposium,
Santa Fe, NM, June 14-21, 1998''}, and references therein.
\end{thebibliography}
\end{document}